\documentclass[final]{aipproc}

\layoutstyle{8x11double}

\begin{document}

\title{Time-scales\,of\,Line-broadening\,Variability\,in\,OB\,Supergiants}


\author{S. Sim\'on-D\'iaz}{
  address={Instituto de Astrof\'isica de Canarias, E38200 La Laguna, Tenerife, Spain}
  ,altaddress={Departamento de Astrofísica, Universidad de La Laguna, E-38205 La Laguna, Tenerife, Spain} 
}

\author{K. Uytterhoeven}{
  address={Laboratoire AIM, CEA/DSM-CNRS-Universit\'e Paris Diderot; CEA, IRFU, SAp, centre de Saclay, 91191, Gif-sur-Yvette, France}
}

\author{A. Herrero}{
  address={Instituto de Astrof\'isica de Canarias, E38200 La Laguna, Tenerife, Spain}
  ,altaddress={Departamento de Astrofísica, Universidad de La Laguna, E-38205 La Laguna, Tenerife, Spain} 
}

\author{N. Castro}{
  address={Instituto de Astrof\'isica de Canarias, E38200 La Laguna, Tenerife, Spain}
  ,altaddress={Departamento de Astrofísica, Universidad de La Laguna, E-38205 La Laguna, Tenerife, Spain} 
}

\begin{abstract}
Several works have recently shown that there is an important extra line-broadening 
(usually called $macroturbulence$) affecting the spectra of O and B Supergiants that 
adds to stellar rotation. So far, the only (very recent) physical explanation for 
the appearance of $macroturbulence$ relates to oscillations. This is a plausible 
explanation, but no direct evidence confirming its validity has been presented yet. 
We recently started an observational project to obtain constraints on the time-scales 
of variability associated to this extra line-broadening and its possible origin.  
Our observational strategy consists of the study of a well selected group of O and B 
stars, for which we obtain time series of high-quality spectra. We present some 
preliminary results from our first campaign with FIES@NOT2.5m.
\end{abstract}

\maketitle

%%%%%%%%%%%%%%%%%%%%%%%%%%%%%%%%%%%%%%%%%%%%
%% MAINMATTER
%%%%%%%%%%%%%%%%%%%%%%%%%%%%%%%%%%%%%%%%%%%%

\section{Introduction}

The presence of an important extra line-broadening (in addition to the rotational 
broadening, and usually called {\em macroturbulence}) affecting the spectra of O 
and B Supergiants (Sgs) has been confirmed by several authors since the first 
studies of linewidth measurements for early-type stars by \cite{Con77}. 
It was initially suggested 
by the deficit of narrow lined objects among these type of stars (\citep{Con77, How97}). 
Later on, the advent of high-quality spectra allowed to confirm that the rotational 
broadening was not enough to fit the line-profiles in some objects, and to investigate 
the possibility to disentangle both broadening contributions. \cite{Rya02} applied a 
{\em goodness-of-fit} method to a sample of high quality spectra of B-Sgs, and obtained 
acceptable results for a model in which the {\em macroturbulence} dominates and rotation is negligible. 
However, the reliability of this method to provide actual values for the rotational and macroturbulent velocities
was somewhat limited by two facts: firstly, the similarity of line-profiles broadened by 
different pairs ($v$\,sin$i$, $v_{\rm macro}$) in $\lambda$-space; in addition, results depend
on the type of broadening profile considered for the {\em macroturbulence} (e.g. gaussian, radial-tangential;
see \cite{Duf06}). \cite{Sim07}, following the ideas proposed by \cite{Gra76}, 
applied the Fourier transform  method to a sample of OB-type stars.  The strength 
of this method is based on the possibility to separate, independently of any other broadening mechanism, the 
$v$sin$i$ of the star from the FT of a line profile. Their study definitely showed that, while the effect of the 
{\em macroturbulence} in OB dwarfs is usually negligible when compared to the rotational broadening, the effect 
of this extra-broadening is clearly present in early type Sgs.\\
\newline

\begin{figure*}
  \includegraphics[height=0.75\textheight, angle=90]{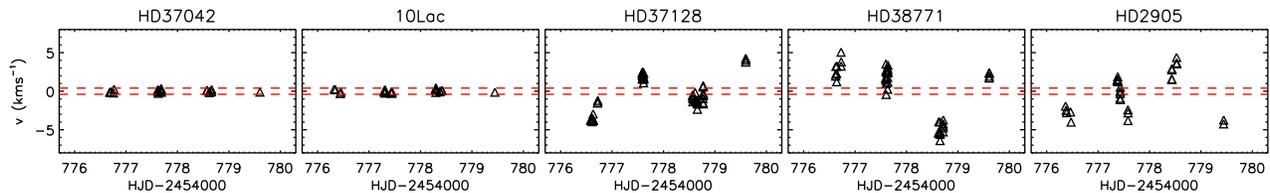}
  \caption{First velocity moment (radial velocity, RV) curves, calculated from the SiIII\,4567 profiles, 
  for the studied stars. 
  Horizontal lines show the dispersion in RVs found for the MS stars. We adopt this values as the accuracy 
  of the RV measurements.}\label{fig:a}
\end{figure*}

Despite it was named {\em macroturbulence} at some point, the interpretation of this extra-broadening as the 
effect of turbulent motions is quite improbable. The effect is present in photospheric lines and affects the whole 
profile, even the wavelengths close to the continuum. Therefore, whatever is producing the extra-broadening has 
to be deeply rooted in the stellar photosphere (and maybe below), in layers in which we do not expect any significant 
velocity field in these stars. However, if interpreted as turbulent motions, {\em macroturbulence} represents 
{\em highly supersonic} velocities in most of the cases (\cite{Duf06}, \cite{Lef07}, \cite{Mar08}). This interpretation seems incompatible with the %%@
previous statement. 

So far, the only (very recent) physical explanation for the appearance of {\em macroturbulence} relates to oscillations. 
\cite{Lef07} studied a sample of periodically variable B-Sgs and suggested, from 
the location of the stars in the ($T_{\rm eff}$, log\,$g$)-diagram, that this variability is due 
to gravity-mode oscillations. Most of their studied stars show photometric variability with main periods 
between 1 and 3 days, with a few stars having shorter (up to 2.5h) or larger periods (up to 25 days). 
Recently, \cite{Aer08} computed time series of line-profiles for 
evolved massive stars broadened by rotation and thousands of low-amplitude non-radial gravity-mode 
oscillations and showed that the resulting profiles could mimic the observed ones. They conclude that 
{\em macroturbulence} is likely a signature of the collective effect of several pulsation modes. This is a 
plausible explanation, but so far no direct evidence confirming its validity has been presented.\\

We recently started an observational project aimed at investigating the possible time-scales of 
variability associated to the $macroturbulent$ broadening. Our observational strategy consists of 
the study of a well selected group of O and B stars (including main sequence stars and supergiants), 
for which we obtain time series of high-resolution, high signal-to-noise spectra. 

\section{Preliminary results from the first observational dataset}

\begin{table}
\begin{tabular}{llcc}
\hline
    \tablehead{1}{l}{b}{Star}
  & \tablehead{1}{c}{b}{SpT \& LC}
  & \tablehead{1}{c}{b}{$v$\,sin$i$ - FT} 
  & \tablehead{1}{c}{b}{$v$\,sin$i$ - FWHM} \\
\hline
HD\,37128  & B0\,Ia    & 45 & 85 \\
HD\,38771  & B0.5\,Ia  & 55 & 95 \\
HD\,2905   & BC0.7\,Ia & 48 & 88 \\
HD\,214680 & O9\,V     & 20 & 37 \\
HD\,37042  & B0.5\,V   & 32 & 34 \\
\hline
\end{tabular}
\caption{List of studied stars and their spectral type and luminosity class. 
Two $v$\,sin$i$ values (in km\,s$^{-1}$),
obtained through the FT and the FWHM methods are indicated (see \cite{Sim07}). 
The comparison between both quantities gives an idea of the amount of
extra line-broadening affecting the line profile.}
\label{tab:a}
\end{table}

In November\,2008, we obtained a first set of spectra with the FIES cross dispersed 
high-resolution echelle spectrograph attached to the NOT2.5\,m telescope at El Roque de los  
Muchachos observatory on La Palma (Islas Canarias, Spain). We used FIES in the medium
resolution mode (R=46000, $\delta\lambda$=0.03 \AA/pix). The entire spectral range 
3700--7300 \AA\ was covered without gaps in a single fixed setting. We obtained time 
series of spectra for six early B-type Sgs during the four awarded nights. 
The sample was completed with two OB main sequence (MS) stars where $macroturbulence$ 
is negligible with respect to rotational broadening. Here, we present results on the
three brightest Sgs and the two MS stars (see Table \ref{tab:a}).

\subsection{Line profile variability}

We found line-profile variability (LPV) signatures in all the studied Sgs, but not in the MS stars. 
Indications of this variability are shown in Figure \ref{fig:a}, where results for the first velocity
moment of the SiIII 4567 line are presented (see \cite{Aer92}, for a definition of the velocity moments
of a line-profile). We also analysed other lines to 
investigate the effects on different line profiles (e.g. CII 4267, OII 4661, OIII 5592, 
and SiIII 4552, among others). Similar results were found for the various considered 
lines.

\subsection{Frequency analysis}

\begin{figure}
  \includegraphics[height=0.24\textheight, angle=-90]{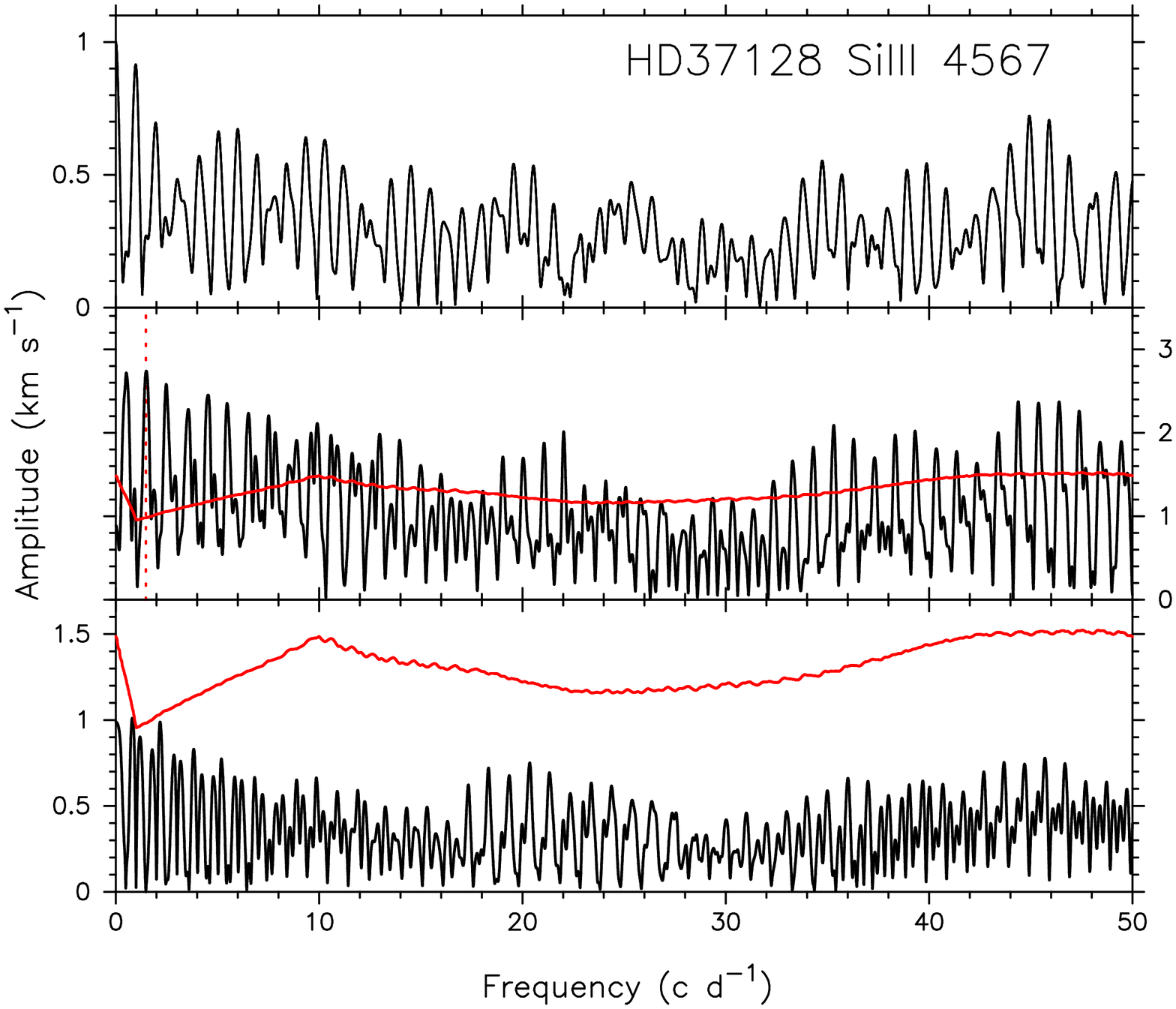}
  \includegraphics[height=0.24\textheight, angle=-90]{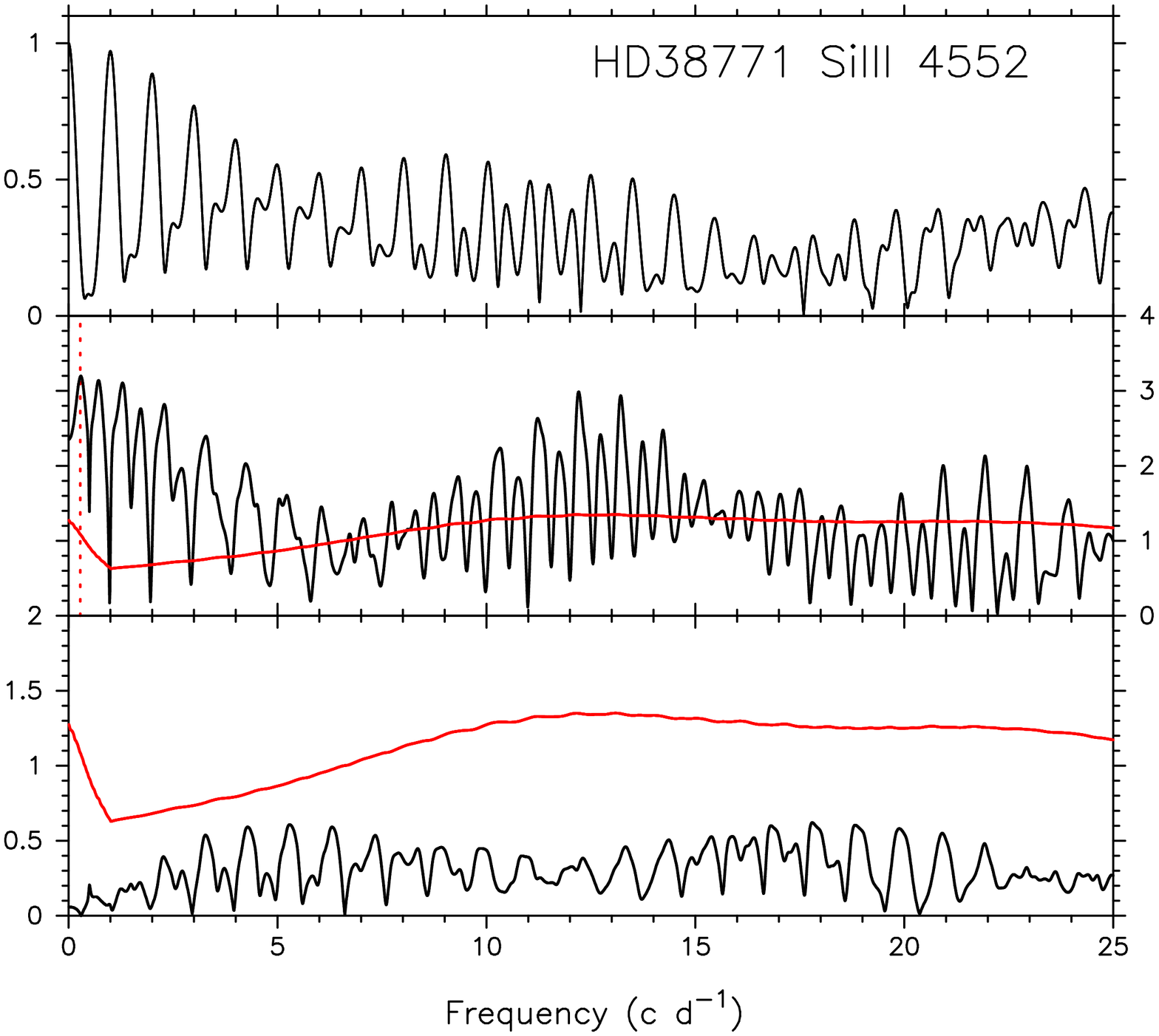}
  \includegraphics[height=0.24\textheight, angle=-90]{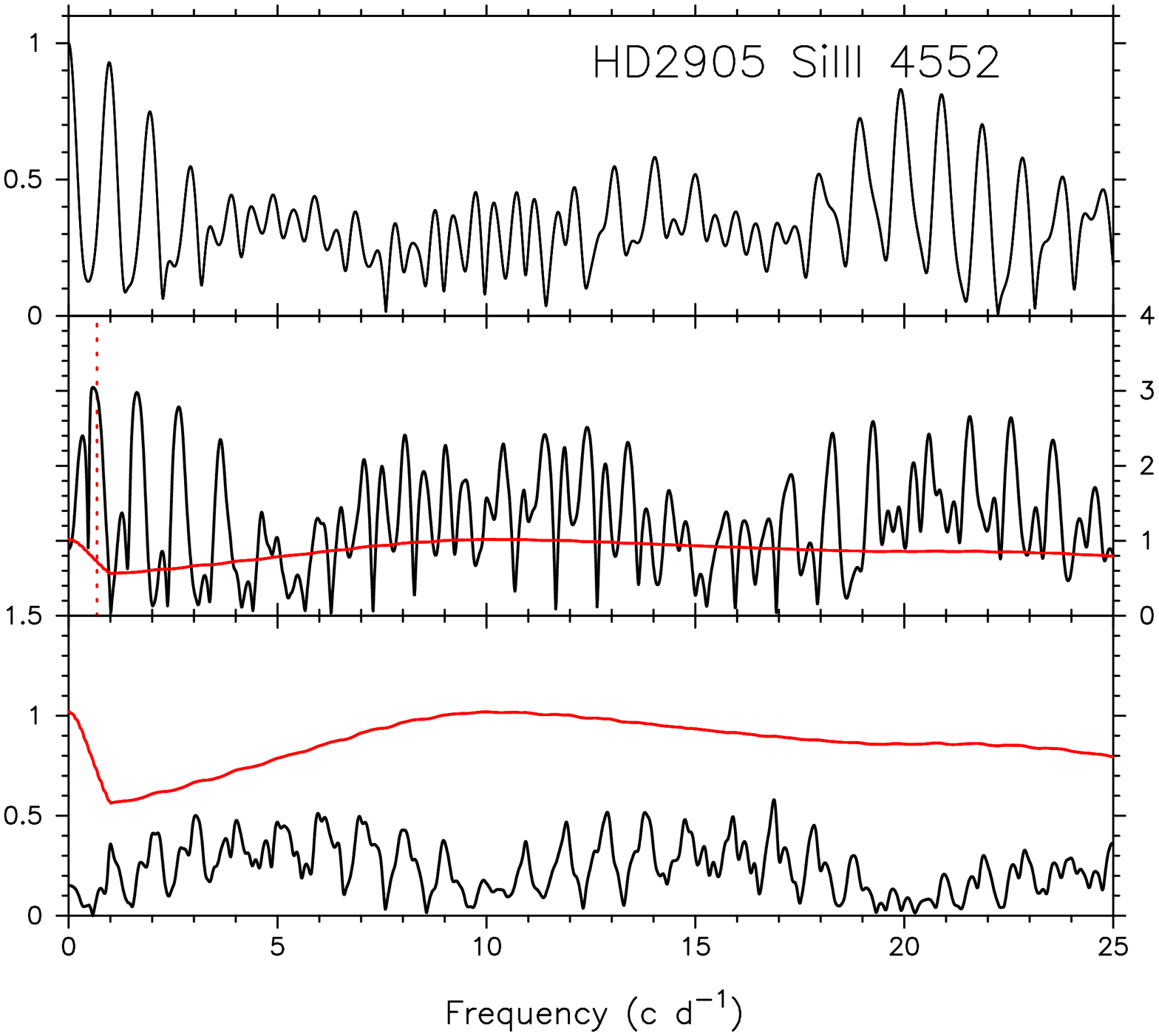}
  \caption{Scargle periodograms calculated from the first moment of the SiIII 4552 line for 
	  the three studies Sgs. Top panels: The associated window function. It is clear that 
	  the data suffer from serious aliasing problems and that the frequency resolution is poor 
	  (~0.02 c/d at best). Middle panels: Amplitude spectrum. Bottom panels: Residual amplitude 
	  spectrum after prewhitening the frequency indicated by the dashed line in the middle panel. 
	  The 4 S/N significance level, calculated as 
	  the average amplitude over a frequency interval of width 2 c/d in the oversampled residual 
	  SCARGLE periodogram, is given by a red solid line.}\label{fig:b}
\end{figure}

We searched for intrinsic frequencies by applying two line-diagnostics: the pixel-to-pixel 
variations across the line profile (IPS method \cite{Tel97}), and the moment variations (\cite{Aer92}).
For the analysis of the moment variations, we use SCARGLE (\cite{Sca82}) and the least-squares 
power spectrum method (\cite{Van71}). We present here only the preliminary results from the 
first velocity moment variations (RV).

\begin{figure}
  \includegraphics[height=0.29\textheight, angle=90]{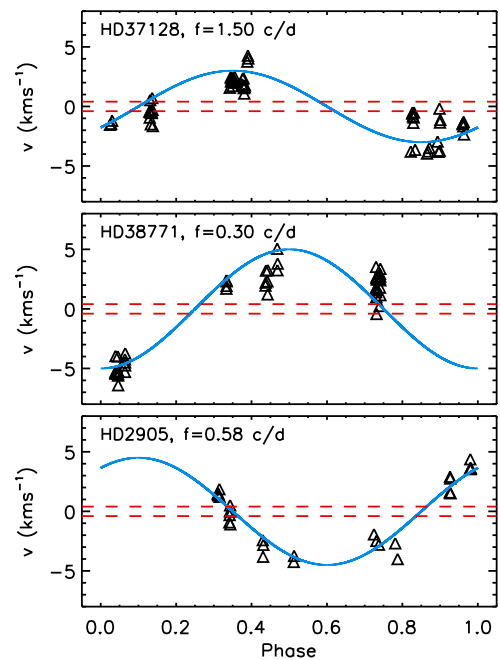}
  \caption{RV curves from the SiIII 4567 line 
folded according to the highest amplitude 
period (see Table \ref{tab:b}). Similarly to Fig. \ref{fig:a}, 
horizontal lines show the dispersion in RV 
found for MS stars. A short-term variability 
is suspected.}\label{fig:c}
\end{figure}

The RV variations suggest that the B-Sgs show variabilities on two time-scales: a variation 
of the order of half a day to several days (Table \ref{tab:b}), with amplitudes of the 
order of 1-9 km/s (also detected from the IPS analysis), and a faster variation of tens of 
minutes at low amplitude (see Fig. \ref{fig:c}). However, only the longer period is found significant in the current 
datasets (4 S/N criterion; \cite{Bre93}, \cite{Kus97}, see Fig. \ref{fig:b}). Although this period is clearly present 
in the LPV of all line profiles investigated, its value cannot be accurately determined  
given the poor frequency resolution (of the order of 0.02 c/d at best) and the influence 
of a bad spectral window (see top panel Fig. \ref{fig:b}). Also, we need more data to investigate the reality of the 
short-term variations. A first comparison of the dispersion in RV between B-Sgs and MS 
stars suggests that the short periods might be intrinsic to the B-Sgs, and not caused by 
instrumental effects (see Figs. \ref{fig:a} and \ref{fig:c}).

\section{Summary and future work}

Our preliminary analysis of line-profile variations in B-Sgs from the first FIES@NOT 
campaign already points to the presence of small-scale periodic variations which are 
not detected in the MS stars. RVs variations show variabilities of the order of half 
a day to several days, with amplitudes of the order of 1-9 km/s. Similar results are 
obtained from lines of different ions/elements. The frequency analysis also seems to 
indicate short-term variations with lower amplitudes. More observations (already planned)
are needed to confirm and improve these results. From the new data we expect to be 
able to constrain the effect of $macroturbulence$ and to test its interpretation in 
terms of stellar pulsations.

\bibliographystyle{aipproc}   % if natbib is available

\begin{table}
\begin{tabular}{lrrc}
\hline
    \tablehead{1}{l}{b}{Star}
  & \tablehead{1}{c}{b}{N}
  & \tablehead{1}{c}{b}{$\Delta$T (d)}
  & \tablehead{1}{c}{b}{Dominant frequency (c/d)}   \\
\hline
HD\,37128 & 60 & 3.02 & 1.50$^*$$\pm$ 1  \\
HD\,38771 & 48 & 2.99 & 0.30$^*$$\pm$ 1  \\
HD\,2905  & 30 & 3.08 & 0.58$^*$$\pm$ 1  \\
\hline
\end{tabular}
\caption{Highest amplitude frequencies resulting from the frequency analysis for 
the B-Sgs. Total number of spectra (N) and the timespan ($\Delta$T) of the 
observations are also indicated. Note: (*) or one of its aliases.}
\label{tab:b}
\end{table}

\end{document}